\documentclass[twocolumn,aps,prb,superscriptaddress,showpacs]{revtex4-1}
\usepackage{graphicx}
\usepackage{amssymb}
\usepackage{color}

\begin{document}

\title{
Geometrical origin
of ferrimagnetism and superparamagnetism in Fe-based double perovskite
multiferroics
}

\author{R.O.\ Kuzian} 
\affiliation{Institute for Problems of Materials Science
NASU, Krzhizhanovskogo 3, 03180 Kiev, Ukraine} 
\affiliation{Donostia International Physics Center (DIPC), ES-20018
Donostia-SanSebastian, Spain}

\author{V.V.\ Laguta} 
\affiliation{Institute for Problems of Materials Science
NASU, Krzhizhanovskogo 3, 03180 Kiev, Ukraine} 
\affiliation{Institute of
Physics, AS CR, Cukrovarnicka 10, 16253 Prague, Czech Republic}

\author{J.\ Richter}
 \affiliation{Institut f\"ur Theoretische Physik,
 Otto-von-Guericke-Universit\"at Magdeburg,\\
PF 4120, D - 39016 Magdeburg, Germany}

\begin{abstract} 
We show that a superstructure of  
antiferromagnetically interacting  Fe$^{3+}$ ($S=5/2$) ions in
double perovskites AFe$_{1/2}$M$_{1/2}$O$_{3}$ exhibits a
ferrimagnetic ordering below $T_{fe} \approx 5.6J_1$ 
($J_1/k_B \sim 50$~K), which is close to room temperature.
Small clusters of the same structure
exhibit a  superparamagnetic behavior at $T \lesssim T_{fe}$. 
The possibility of formation of such clusters explains 
the room-temperature 
(superpara)magnetism in 3$d$-metal based oxides.
\end{abstract}

\date{16.01.14} 


\maketitle 

\section{Introduction}
An experimental quest to find a room-temperature 
multiferroic with high magnetoelectric coupling  is 
stimulated by
wide prospects they open for applications in the field of information 
and
energy-saving technologies. They may form the basis for a fabrication of 
novel
functional devices: highly sensitive magnetic sensors, 
capacitance electromagnets, 
elements of
magnetic memory switched by electric field, 
nonreciprocal microwave filters, and others.\cite{Pyatakov12,Scott12}
Spintronics, an emerging branch of 
micro- and nanoelectronics which manipulates the electron spin rather 
than 
its charge, has need for a room-temperature ferromagnetic 
semiconductor.\cite{Zutic04}
\begin{figure*}[htb] 
\includegraphics[width= 0.48\textwidth]{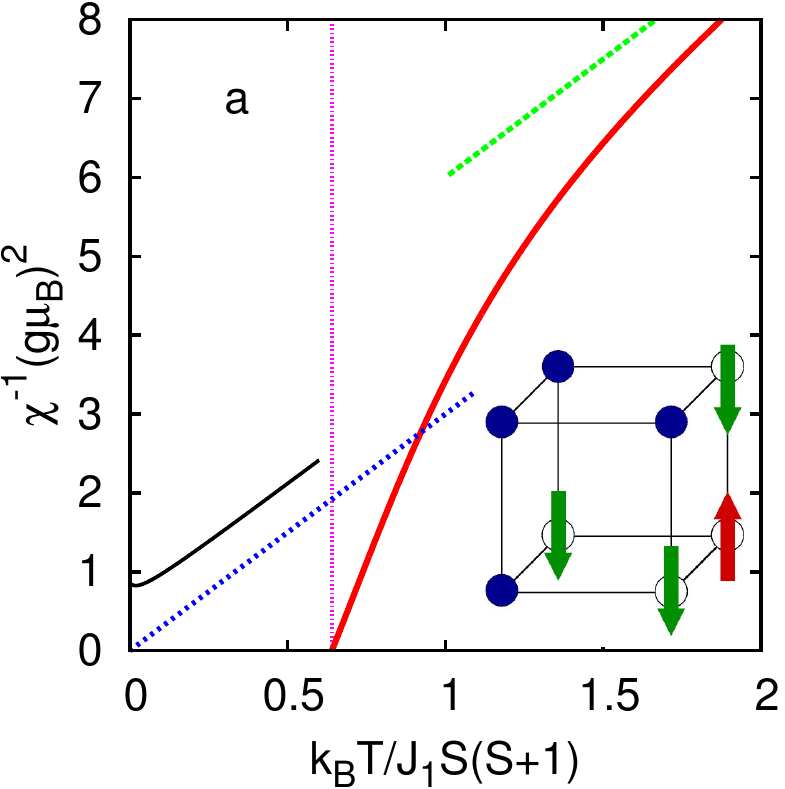} %\\
\includegraphics[width= 0.48\textwidth]{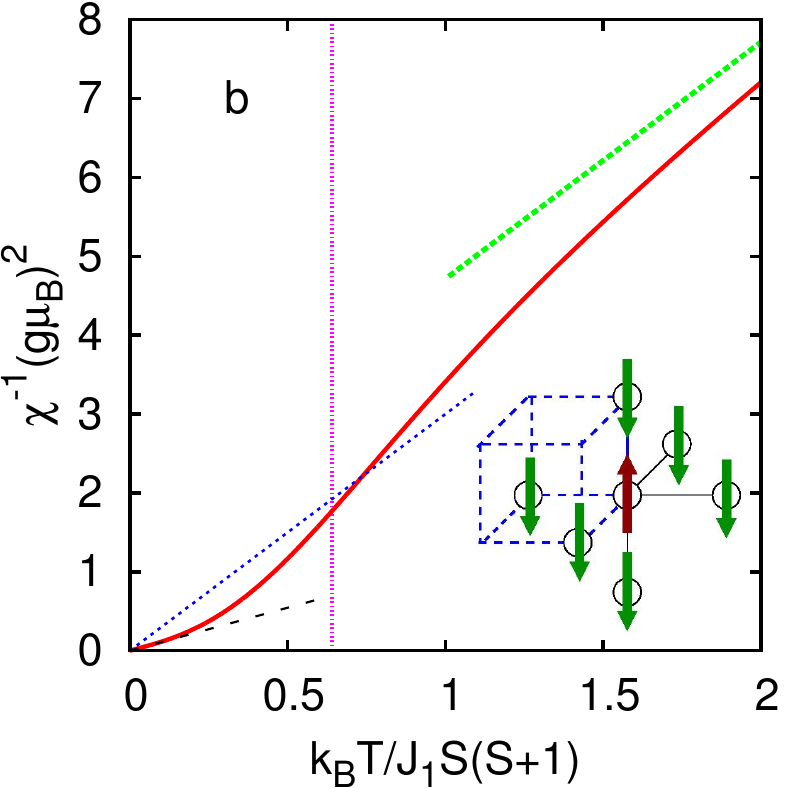}
\caption{(Color online)
\textbf{a}: Inverse subsceptibility $\chi^{-1}$ for a
periodic arrangement of PFB2 chemical order with two 
inequivalent $S=5/2$
Fe$^{3+}$ ion positions (red solid line - 
[4,4] Pad\'e approximant of the 8th order
HTE series).  The susceptibility 
exeeds the Curie-Weiss (CW) asymptotic 
(green dashed line) 
and diverges at $T_{fe} \approx  0.640J_1S(S+1)$ (shown by the 
vertical line) 
corresponding 
to a transition into a ferrimagnetic phase.
The black thin solid line shows the susceptibility 
of 1:1 ordered PFB0 configuration, where Fe spins 
interact with $J_2=0.05J_1$.  
Inset: Unit cell of the PFB2
chemical ordering, only 
Fe (open circles) and M (filled circles) cations
are shown. Fe1(Fe2) positions are depicted by 
up(down) arrows, respectively. 
\textbf{b}: 
Inverse subsceptibility $\chi^{-1}$ (red solid line) for 
a small cluster 
of a PFB2 configuration (as shown in the inset) obtained by full exact 
diagonalization. 
The %inverse 
susceptibility shows a crossover between CW (dashed 
green) and superparamagnetic (dashed black) behavior.
In both parts, the blue dotted line shows the 
Curie law for independent spins, $\chi _{\mathrm{p}}^{-1}
\propto T$.
} 
\label{hte} 
\end{figure*}

The rich family of Fe-based double perovskites
AFe$_{1/2}$M$_{1/2}$O$_{3}$=A$_2$FeMO$_6$ 
(with non-magnetic ions A=Pb,Ca,Sr,Ba, and M=Nb,Ta,Sb) is in the focus of the
studies as it includes
PbFe$_{1/2}$Nb$_{1/2}$O$_{3}$(PFN) 
and PbFe$_{1/2}$Ta$_{1/2}$O$_{3}$ (PFT)
systems, where the multiferroicity was reported more then fifty years
ago.\cite{Smolenskii58,Bokov62}

In AFe$_{1/2}$M$_{1/2}$O$_3$ compositions, Fe$^{3+}$ and M$^{5+}$ 
cation positions may be ordered or disordered  within simple 
cubic B-sublattice of perovskite 
structure ABO$_3$. The degree of chemical ordering  
depends on the strength of electrostatic and 
elastic energies and, in particular, on the ionic radii of these cations. 
It is commonly accepted that PFN and PFT are chemically disordered 
compounds due to almost equal 
ionic radii of Fe$^{3+}$
and Nb$^{5+}$ or Ta$^{5+}$,\cite{Shannon76}
while Sb-contained compounds can be
chemically ordered up to 90\% 
because Sb$^{5+}$ is much larger than Fe$^{3+}$.\cite{Misjul13}
Magnetism of the compositions is due to Fe$^{3+}$, $S=5/2$ ions
that occupy half of octahedral sites of the perovskite lattice.
The magnetic moments of the Fe$^{3+}$ ions
interact via various superexchange
paths, 
\begin{equation}
\hat{H}=\frac{1}{2}\sum_{\mathbf{R,r}}J_{\mathbf{r}}
\hat{\mathbf{S}}_{\mathbf{R}}\hat{\mathbf{S}}_{\mathbf{R+r}}. 
\label{Heis}\end{equation}
The disorder prevents an experimental access to the values of the 
interactions.
In a recent publication, some of us have argued that the 
largest superexchange values are the
nearest-neighbor (NN) Fe-Fe interaction (Fe ions are 
separated by the edge of perovskite unit cell and interact via the 
shortest Fe-O-Fe path) $J_1 \sim 50 - 70$~K and the next-nearest-neighbor interaction 
(Fe ions are separated by the face
diagonal of the cell) $J_2\simeq 0.04J_1$.\cite{Kuzian14}
The interaction values $J_1$, $J_2$ are similar to the values in
orthoferrite RFeO$_3$ (R=Y or a rare
earth) \cite{Gorodetsky69,Shapiro74,Gukasov97,Delaire12,McQueeney08} 
and bismuth ferrite BiFeO$_3$ \cite{Gabbasova91}
compounds. Note that both exchange couplings 
have \emph{antiferromagnetic} sign. 
We thus have two substantially different magnetic energy scales:
$S(S+1)J_{1}=8.75J_{1}$, which corresponds to temperatures of several 
hundred Kelvins, and $S(S+1)J_{2}/k_B \sim 20$~K. 
Note that many of Fe-based double
perovskites have an antiferromagnetic phase transition in the latter temperature
range.\cite{Battle95,Battle95a,Tezuka00,Kleemann10,
Raevski09,Laguta13} It
means that  the probability to find a pair of Fe ions 
separated by the face
diagonal of the perovskite cell is much higher than to 
find a nearest-neighbor
Fe pair that is caused by partial chemical ordering of cations. 
Two multiferroic compounds, PFN and PFT, 
exhibit a magnetic transition at $T_N
\sim 150$~K.
%, which belongs to the large energy scale $\sim S(S+1)J_{1}$. 
This
means that the probability to find a pair of NN Fe ions is enhanced in 
these compounds.
%, compared to the other double perovskites. 
But it leads to the
increase of the temperature, at which the \emph{antiferromagnetic} order is
established.\cite{Pietrzak81,Ivanov00,Rotaru09} For instance, in
the more 
concentrated compound PbFe$_{2/3}$W$_{1/3}$O$_3$ it increases up 
to 380~K.\cite{Ivanov04}

Recent reports on room-temperature multiferroicity of PFT/lead zirconate
titanate (PZT) \cite{Sanchez11,Evans13} and PFN/PZT\cite{Sanchez13} 
and [Pb(Fe$_{2/3}$W$_{1/3}$)O$_3$]/PZT \cite{Kumar09} solid solution
systems are a real challenge for the solid state theory. 
One of the questions is the nature of
large room-temperature magnetic response of the systems
(non-linear magnetization curves and hysteresis loops) that 
imply the existence of Fe spins alignment  in a part of the sample
with uncompensated magnetic moment.
On the qualitative level, it was suggested 
that the clustering of Fe ions is responsible for 
the appearence of the uncompensated magnetic 
moment.\cite{Blinc07,Kumar09,Sanchez11,Evans13,Sanchez13}
We should mention that
the clustering of Fe ions \cite{Laguta10,Raevski12,*raevtab}
forms locally fragments of 
AFeO$_3$ structure, where Fe spins form
the simple cubic lattice. Thus, it
can lead only to G-type antiferromagnetic ordering 
within the fragments, and
produces a small or vanishing uncompensated magnetic moment. It
\emph{can not} convincingly 
explain the observation of
room-temperature hysteresis loops.

A small canting of predominantly antiferromagnetic Fe spins
due to the antisymmetric Dzyaloshinskii-Moriya interaction 
$\hat{H}_{DM}=\mathbf{D}\cdot[\mathbf{S}_1 \times \mathbf{S}_2]$
causes weak ferromagnetism 
in ortoferrites RFeO$_3$, R$^{3+}$ being Y or a rare earth
ion. It was suggested that the canting may cause also the uncompensated
magnetic moment in AFeO$_3$ structure  that is formed by the Fe ions 
clustering in the double
perovskites.\cite{Majumder06,Sanchez11}
But the moment
seems to be too small to explain the
effect.\cite{Keffer62,Moskvin75ru,*Moskvin75en}
In the ordered state of RFeO$_3$, the canting angle 
$\phi \sim 10$~mrad results in the moment
$\sigma \sim 0.05\mu _B$ per Fe ion.\cite{Treves65,Lutgemeier80}
But such a moment 
was never observed in the antiferromagnetically ordered state of PFN
neither in magnetic \cite{Kleemann10,Laguta13} nor in neutron 
\cite{Pietrzak81,Ivanov00,Rotaru09} studies. A possible reason is 
that the Dzyaloshinskii-Moriya vector for a Fe-O-Fe bond
may be written as \cite{Keffer62,Moskvin75ru,*Moskvin75en}
$\mathbf{D}=d[\mathbf{r}_1\times \mathbf{r}_2]$,
where $d$ is a scalar value, and $\mathbf{r}_i$ is a unit vector in 
the direction from oxygen to spin $\mathbf{S}_i$. Thus, its value 
depends on the Fe-O-Fe bond angle 
$D\propto \sin \theta$, 
which is substantially larger in AFe$_{1/2}$M$_{1/2}$O$_{3}$  
($170^{\circ} <\theta < 180^{\circ} $)\cite{Ivanov00,Misjul13}
than in the orthoferrites  ($140< \theta < 157$) \cite{Treves65pl}.

In this paper, 
we quantitatively consider another scenario for 
the room-temperature magnetism of bulk
PFT/PZT and PFN/PZT systems
\cite{Sanchez11,Evans13,Sanchez13} and superparamagnetism often 
observed in PFN nanoparticles  
or even ceramics and thin films.\cite{Blinc07,Correa08,Peng09} 
We explain it by the existence of regions
with a special chemical order 
(a sub-nano-size superstructure)
that results in a 
{\em ferrimagnetic} ordering of
antiferromagnetically interacting Fe$^{3+}$ $S=5/2$ spins.
This explanation was implicitly assumed in Ref. \onlinecite{Blinc07}, 
where the observed slightly asymmetric EPR line shapes above room 
temperature were simulated by a model 
involving the presence of thermally fluctuating 
superparamagneticlike nanoclusters. 
Note that our explanation \emph{does not} demand the clusterization, 
as the stoichiometry 
AFe$_{1/2}$M$_{1/2}$O$_{3}$ is retained within the 
$2 \times 2 \times 2$ supercell of the superstructure.
Using the 
%eighth-order 
high-temperature expansion (HTE),\cite{Schmidt11,*hte,Lohmann14}
we show that a
macroscopic number of spins orders at about the room temperature, 
whereas small
clusters (studied by exact diagonalization method)
exhibit a crossover between paramagnetic and superparamagnetic
behavior.
\begin{figure}[htb]
\includegraphics[width=0.2\columnwidth]{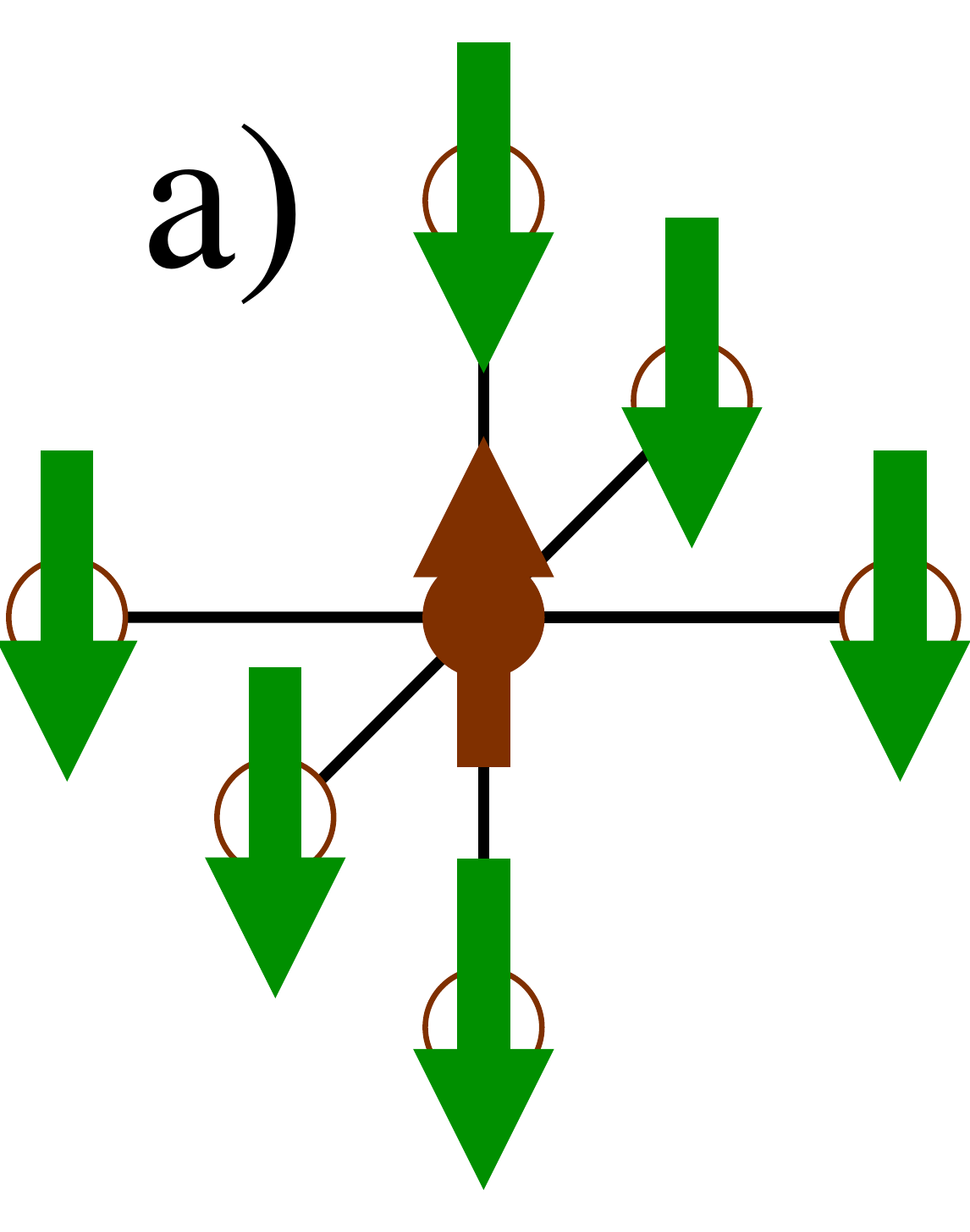}\quad
\includegraphics[width=0.37\columnwidth]{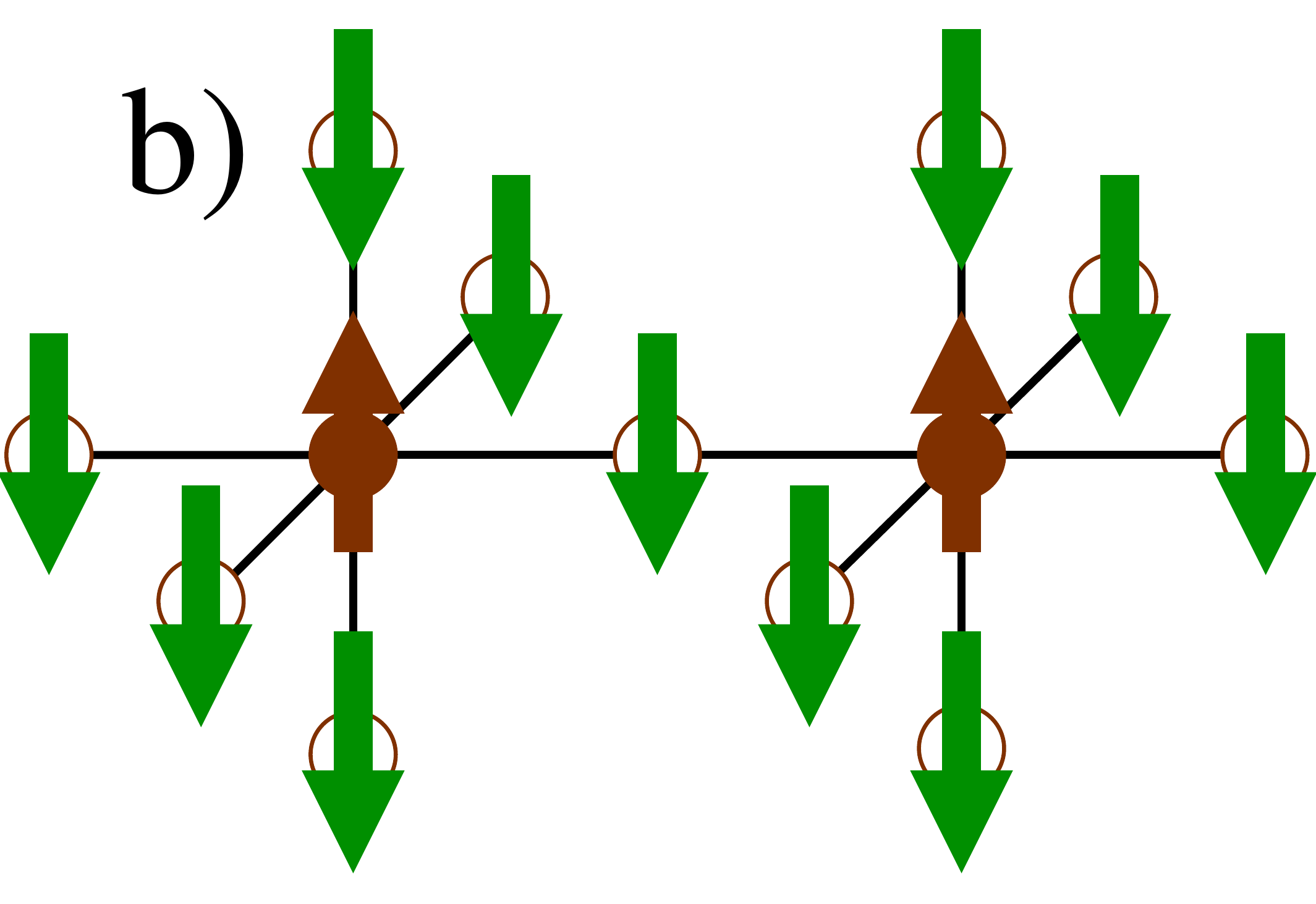}\\
\includegraphics[width=0.5\columnwidth]{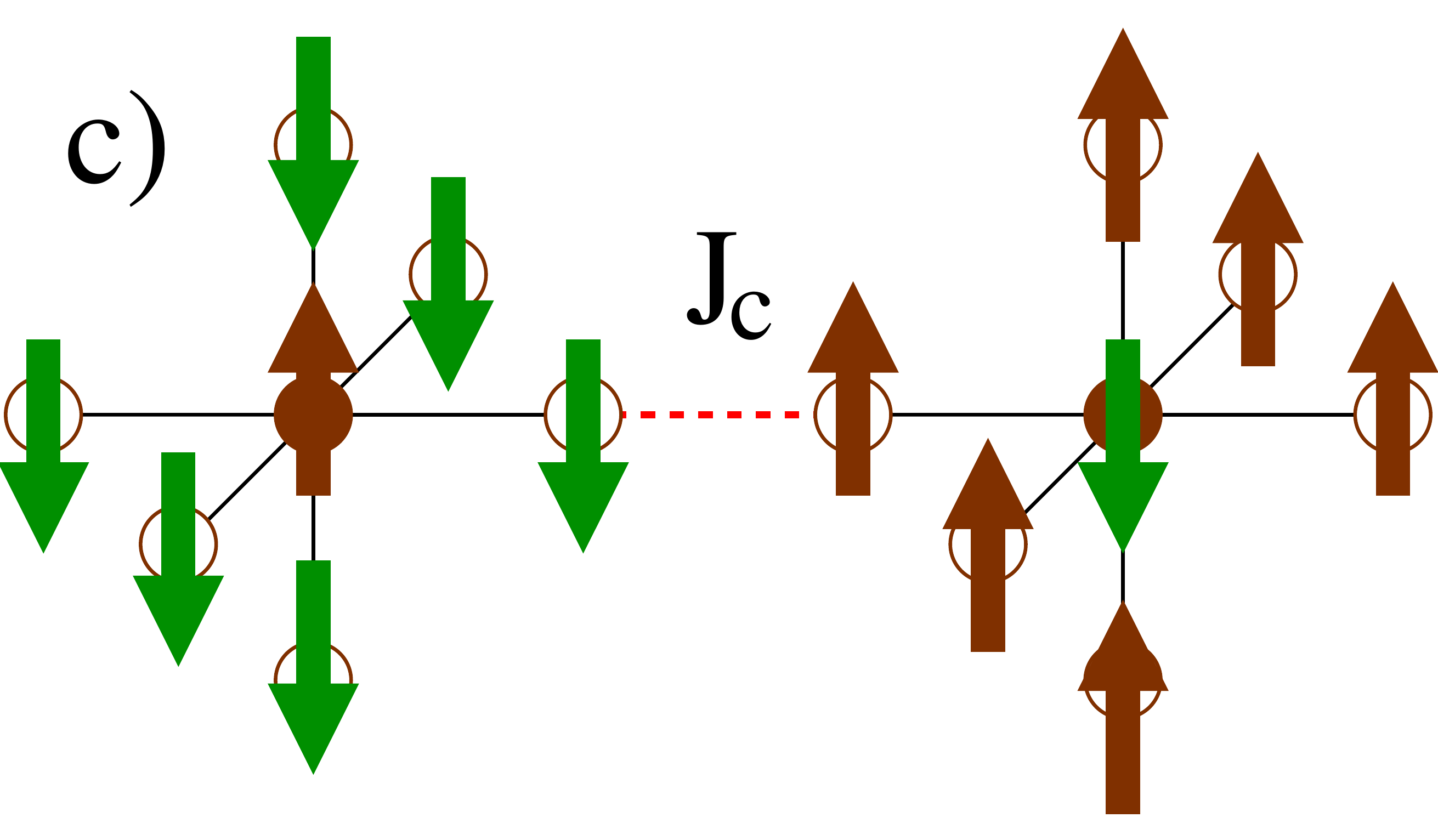}\\
 \includegraphics[width=\columnwidth]{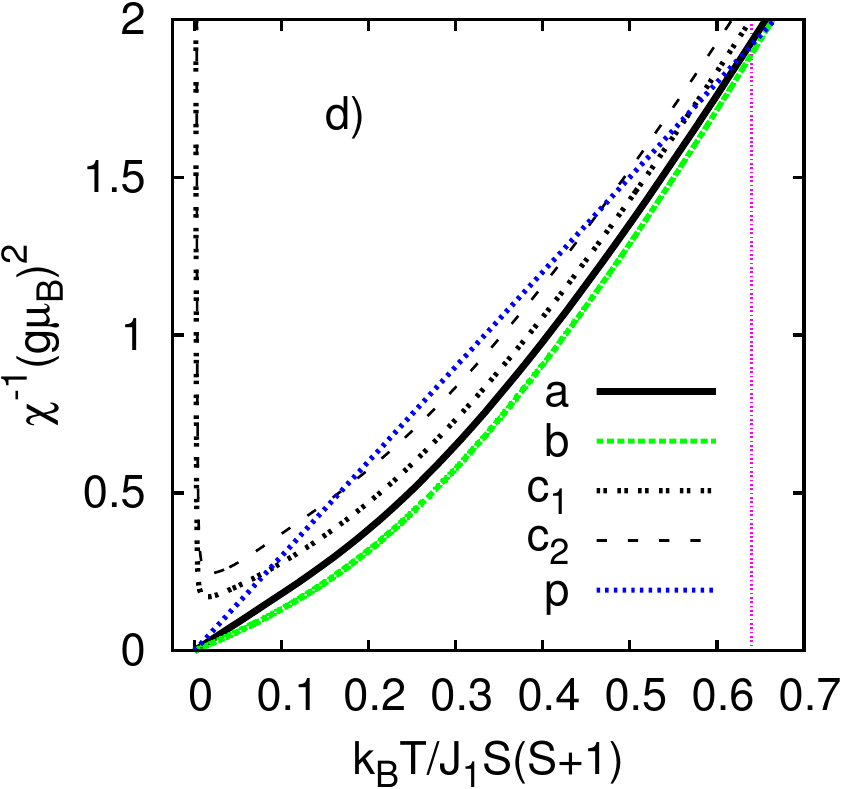}
\caption{(Color online)
(\textbf{a},\textbf{b}) The fragments of PFB2 configuration.
(\textbf{c}) The model simulating two
interacting clusters of PFB2 configuration. 
The coupling strength $J_1$($J_c$) is
denoted by black solid (red dashed) lines. 
Arrows indicate spin-spin correlations in the ground state of each 
cluster.
(\textbf{d}) 
The  temperature dependence of the inverse susceptibility of all 
clusters (full exact diagonalization data for $S=1/2$).
The vertical line shows 
$T_{fe}$. At $T \lesssim T_{fe}$ the susceptibility of the clusters 
exceeds the susceptibility of independent spins (line $p$)
down to the lowest temperatures. The lines $a$ and $b$ 
correspond to clusters with 7 spins \textbf{a} and 13 spins \textbf{b} respectively.
The  lines $c_1$ and $c_2$ correspond to 14-spin cluster \textbf{c}
with $J_c = 0.25J_1$ and $0.5J_1$ respectively.
}
\label{feoct}
\end{figure}

\section{Methods}
We use the method and the program packages presented earlier in the 
Refs.\ \onlinecite{Schmidt11,hte,Lohmann14} for the 
eighth- and tenth-order high-temperature 
expansion (HTE) of the magnetic susceptibility $\chi$
 for a general Heisenberg model with up to four 
different exchange parameters $J_1, J_2, J_3, J_4$.
The input for the HTE package is the definition file where all bonds  
within a cluster or a $L\times L\times L$ (L=16,20) super-cell of a periodic 
Heisenberg lattice are enumerated with the indication of a corresponding 
value of the exchange interaction. We use an
originally developed C++ program, for the generation of the definition 
files for spin structures studied in this work.

In order to simulate the behavior of fragments of PFB2 configuration
in the Fe-based double perovskite material, 
we have performed full exact diagonalization studies (ED) of 
thermodynamic properties of clusters shown in Fig.~\ref{hte}b,
and in Fig.~\ref{feoct}
using J. Schulenburg's {\it spinpack}. 
The susceptibility $\chi (T)$ is calculated as
the ratio of the induced magnetization $M$ to the field $H$.
We use the  "vanishing" magnetic field $H=10^{-5}J_1/g\mu _B$
unless otherwise noted.

\section{Result and Discussion}
\subsection{Ferrimagnetic superstructure}
The simplest way to model the (partial)
disorder in the distribution of Fe and M ions between the sites of the
B-sublattice of the perovskite structure is to consider a 
periodic lattice with a supercell containing
several perovskite cells and study such periodic 
systems with different versions
of chemical order (ion distributions). Such an approach was suggested in
Ref.~\onlinecite{Raevski12} for a $2\times 2\times 2$ supercell, 
where 6 configurations PFB0\dots PFB5 (see Fig.~3 of 
Ref.~\onlinecite{Raevski12}, and Fig.\ 2 of Ref.\ \onlinecite{Kuzian14}) 
of chemical ordering are possible in the double perovskites. 
It was shown that the total energy is
substantially different for different configurations.
Moreover,
the hierarchy of the energies depends on the type of M-ion.
In Ref.\ \onlinecite{Kuzian14}, it was found that
the PFB2 configuration shown in the inset  of Fig.~\ref{hte}a
has an energy close to the most stable  configurations
(PFB5 for M=Nb,Ta and PFB0 for M=Sb), and
has a \emph{ferrimagnetic} ground state (see Table II of 
Ref.\ \onlinecite{Kuzian14}).
Below, we consider the ferrimagnetism of PFB2 superstructure 
in more detail. 

The PFB2 chemical order has two inequivalent Fe sites. 
Within the B-sublattice of the perovskite structure, Fe1 has six Fe2 NN ions, 
whereas three Fe2 sites in the supercell has only two Fe1 NN ions
(insets in Fig. \ref{hte}a,b). In other words, Fe2 sites form a 
superstructure of corner-shared octahedra, Fe1 sites being in the 
center of each octahedron. 
The interaction value between the two
sublattices is $J_1$, and within Fe2 sublattice is $J_2 \ll J_1$.
Thus, the spin system satisfies the requirements of the Lieb-Mattis 
theorem \cite{Lieb62} with $g_{LM}^2 = J_2/4$ 
(see Eq.(2) of the Ref. \onlinecite{Lieb62}). Moreover, it is close to 
the special case $g_{LM}^2 = 0$.
According to the theorem
(see also the consideration of frustration $J_2 \neq 0$ in the
Ref.\ \onlinecite{Richter95}),
the PFB2 ground state corresponds to a \emph{ferrimagnetic} ordering 
of Fe spins with a magnetic moment
of $2g\mu _BS \approx 10\mu _B$ per supercell, or $2.5\mu _B$
per Fe ion. This moment value is much larger than the value provided 
by Dzyaloshinskii-Moriya interaction for realistic values of 
local lattice distortions.\cite{Keffer62,Moskvin75ru,*Moskvin75en,Lutgemeier80}
 
We use the [4,4] Pad\'e
approximant of the HTE series to analyse the susceptibilty 
data.\cite{Schmidt11,*hte}
For a magnetic superstructure with the PFB2 spin arrangement the
temperature dependence of the inverse susceptibility $\chi^{-1}(T)$ 
for $S=5/2$ is shown  in the Fig.~\ref{hte}a. 
Only NN interaction $J_1\neq 0$ was taken into account.
A reasonable estimate of the temperature for the transition into
the ferrimagnetically ordered phase 
$T_{fe}$ is given by that point where 
$\chi ^{-1}(T_{fe})=0$. 
The  precision of the determination of critical temperatures 
by the zero of $\chi ^{-1}$ was estimated to be about 10\%.\cite{Lohmann14}
The values of $T_{fe}$ for different spin values are given in 
the Table \ref{tab:Tfe}. 
\begin{table}[htb]
\caption{\label{tab:Tfe}
The ferrimagnetic transition temperature for PFB2 configuration, 
obtained from the [4,4] Pad\'e
approximant of the 8th order HTE series.
}
\begin{ruledtabular}  \begin{tabular}{ccc}
Spin, $S$ & $k_BT_{fe}/J_1S(S+1)$ & $k_BT_{fe}/J_1$ \\
\hline
1/2 & 0.61 & 0.46\\
1 & 0.69 & 1.4\\
3/2 & 0.64 & 2.4\\
2 & 0.64 & 3.8\\
5/2 & 0.64 & 5.6
\end{tabular}\end{ruledtabular}
\end{table}
\begin{figure}[htb]
\includegraphics[width=\columnwidth]{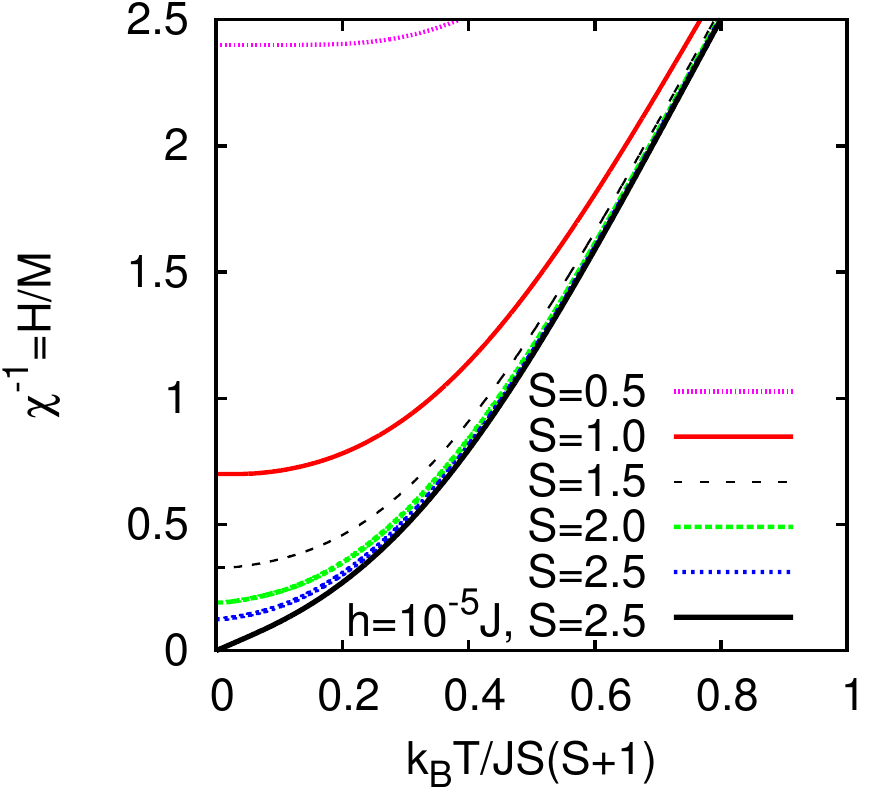}
\caption{(Color online)
The inverse susceptibility $\chi ^{-1}(T)=H/M$
for the cluster shown in the Fig.~\ref{feoct}a
and the field $H=3J_1/[(5S+1)g\mu _B]$
for different spin values. For comparison, an $S=5/2$ curve for a 
vanishing field, i.e.
$\chi ^{-1} \approx (\partial M/\partial H)^{-1}(H=0)$,
is given by the black solid line.
}
\label{hvsm}
\end{figure}
\begin{figure}[htb] 
\includegraphics[width= 0.45\columnwidth]{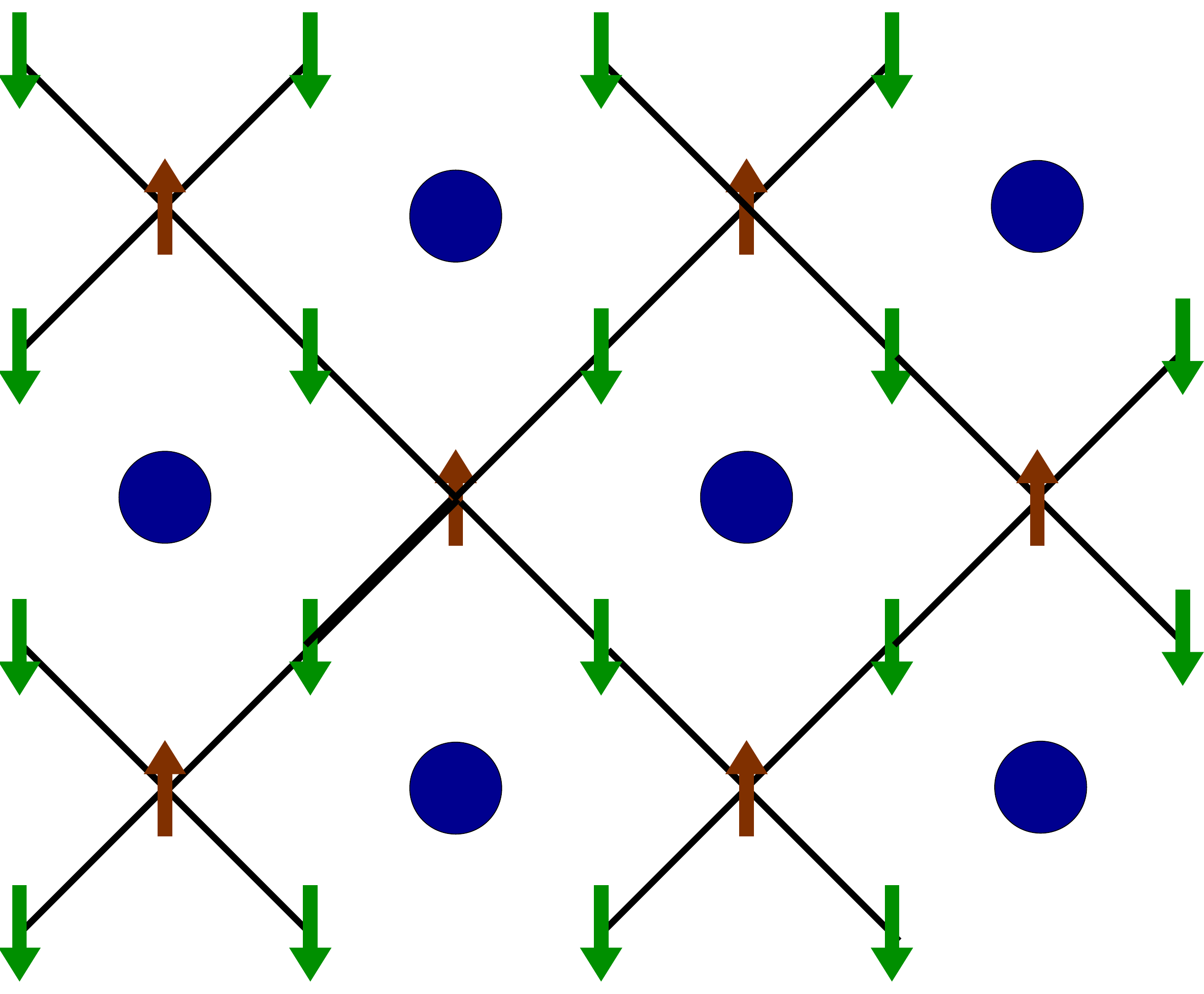}
\includegraphics[width= 0.5\columnwidth]{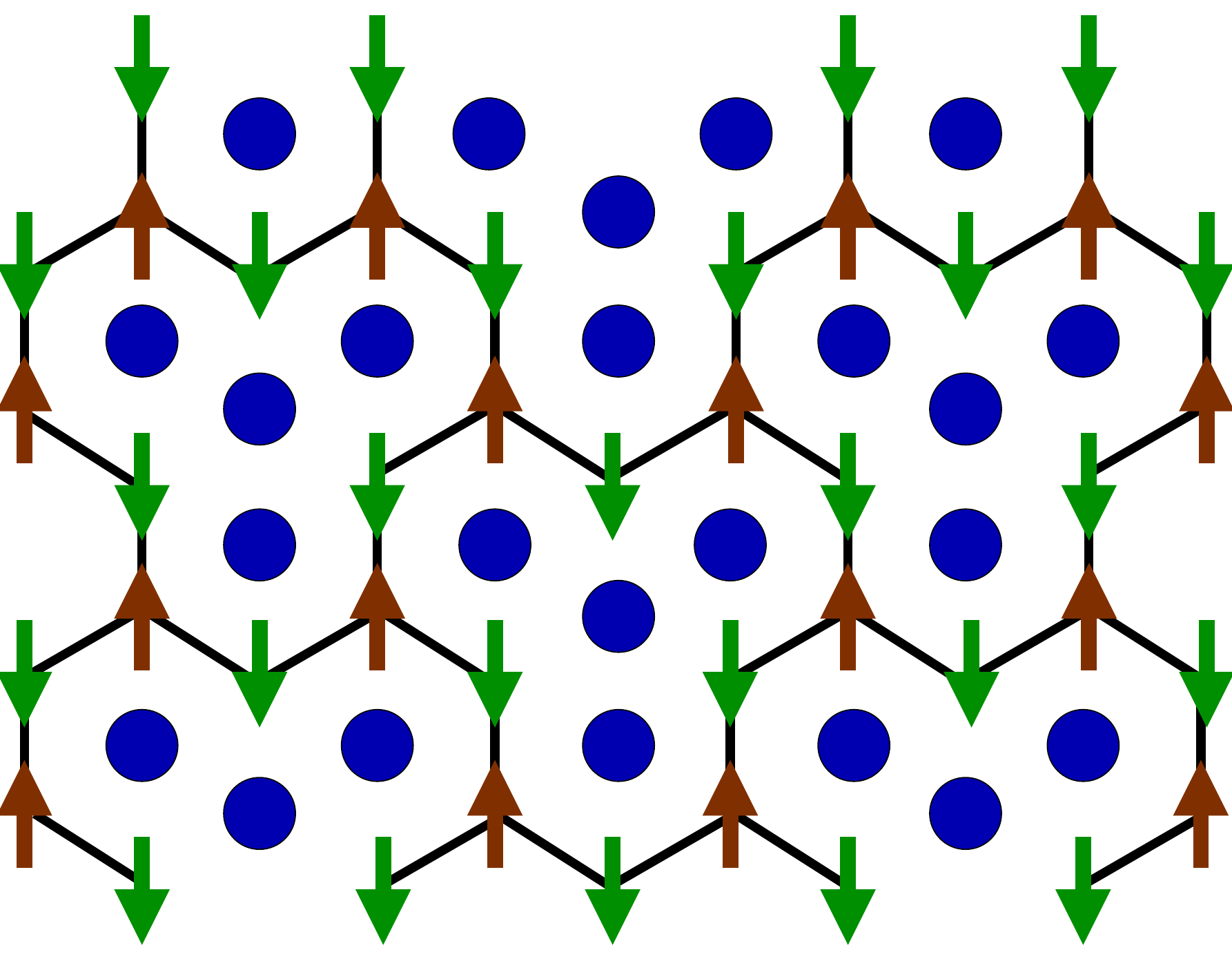}
\caption{(Color online)
Examples of ferrimagnetic superstructures that may be formed by
magnetic impurities (arrows) substituting for cations (blue circles) 
in zinc blend (left)
and wurtzit (right) lattices.
}
\end{figure}
\begin{figure}[htb]
\includegraphics[width=0.8\columnwidth]{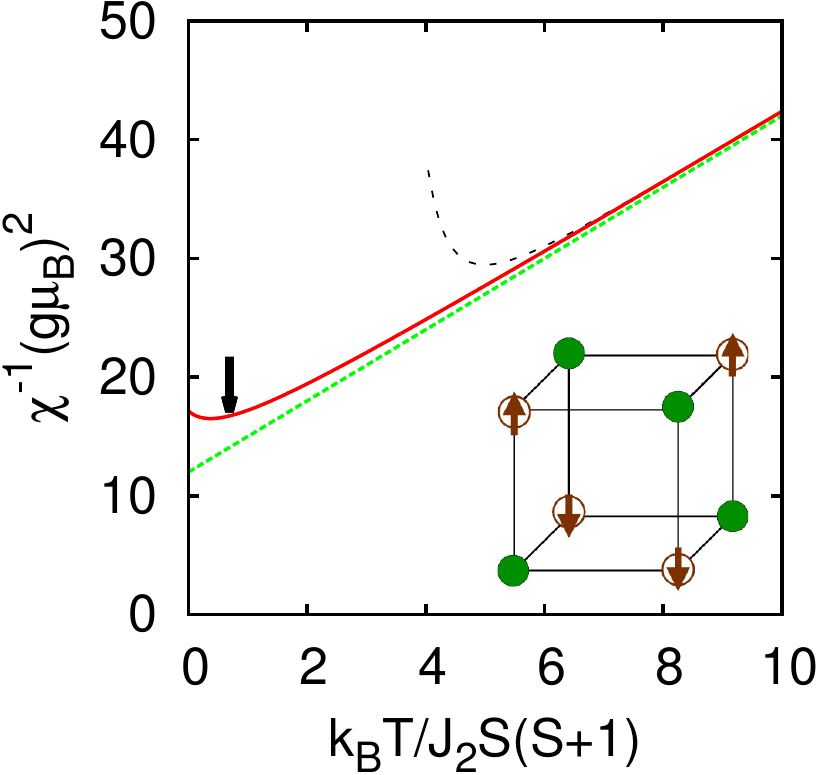}
\caption{Main panel: The inverse magnetic susceptibility (per
spin) $\chi^{-1}$ (red solid line - 
[4,4] Pad\'e approximant of the 8th order
HTE series) for the  ideal 1:1 chemical order (PFB0, shown
in the insert). It shows a minimum at $T\sim T_I$ idicated by the arrow.
The Curie-Weiss  asymptotic is shown by the green dotted
 line. Black dotted line shows the bare HTE series.
Inset: The super-cell for the PFB0, only M$^{5+}$ (closed circles) and
Fe$^{3+}$ ions (open circles) are shown. Arrows indicate  
the distribution of spins in the I-type ordering.}
\label{pfb0}
\end{figure}
\begin{figure}[htb]
\includegraphics[width=0.8\columnwidth]{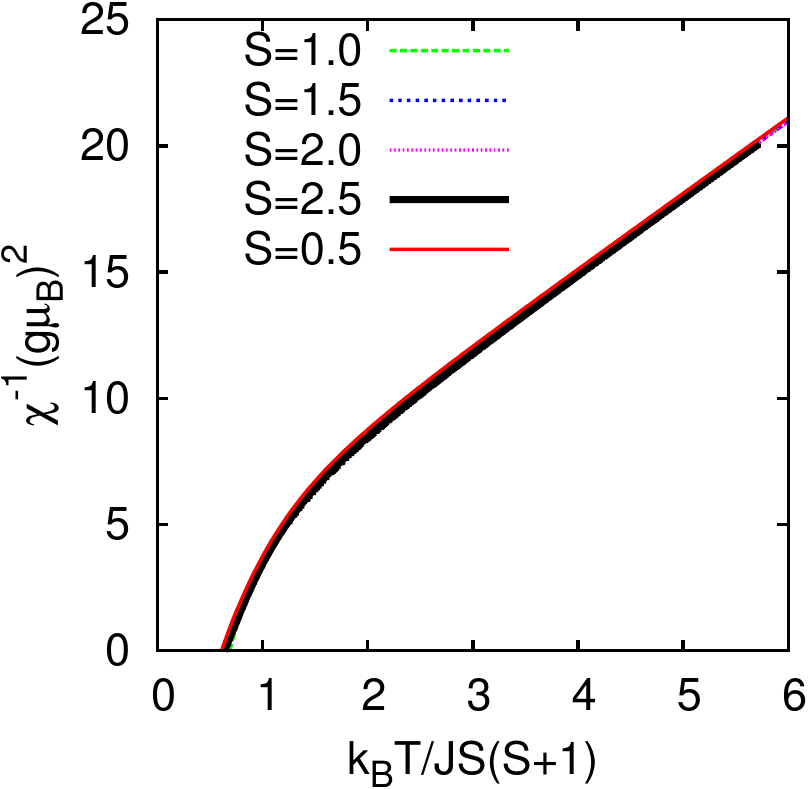}
\includegraphics[width=0.8\columnwidth]{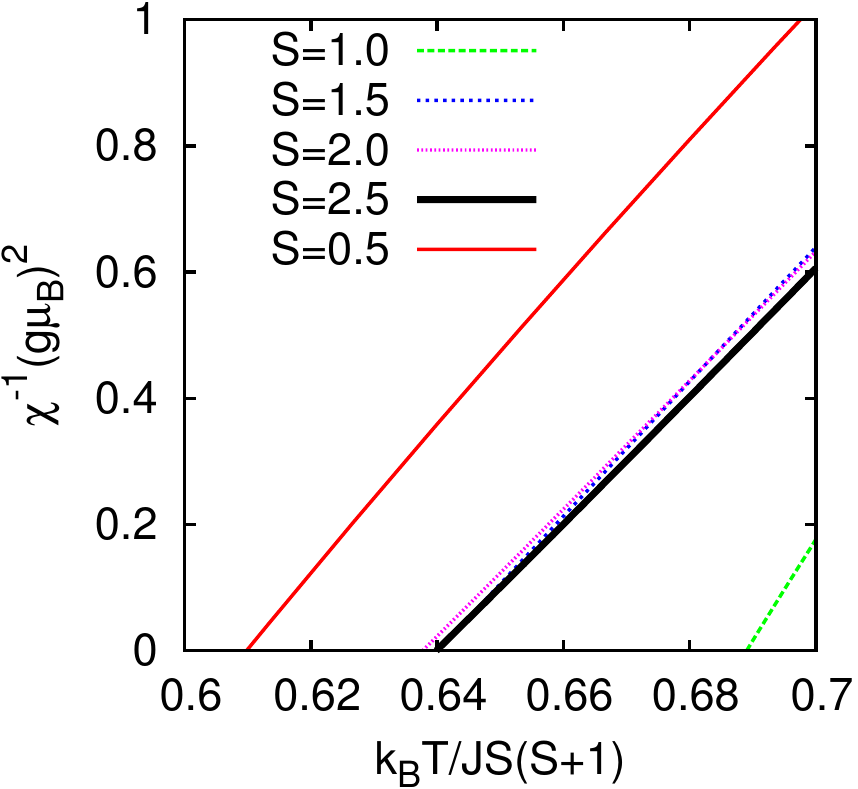}
\caption{Temperature dependence of inverse magnetic susceptibility per 
spin for PFB2 chemical order for systems with different spin
values $S$. [4,4] Pad\'e approximant of the 8th order
HTE series are shown for $S > 0.5$, and [4,6] Pad\'e approximant of the 
10th order HTE series for $S=0.5$
}
\label{pfb2}
\end{figure}
\begin{figure}[b]
\includegraphics[width=0.8\columnwidth]{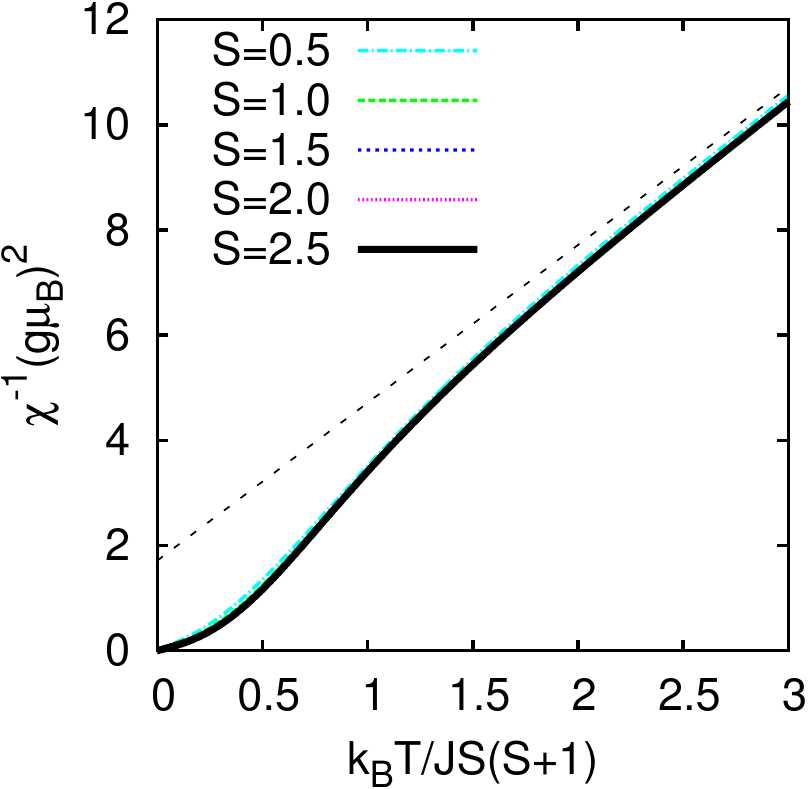}
\caption{Temperature dependence of inverse magnetic susceptibility per 
spin for the cluster shown in Fig.\ \ref{feoct}a and various spin 
values. Dotted line shows Curie-Weiss asymptotic. 
}
\label{feoctc}
\end{figure}
\begin{figure}[htb]
\includegraphics[width=0.8\columnwidth]{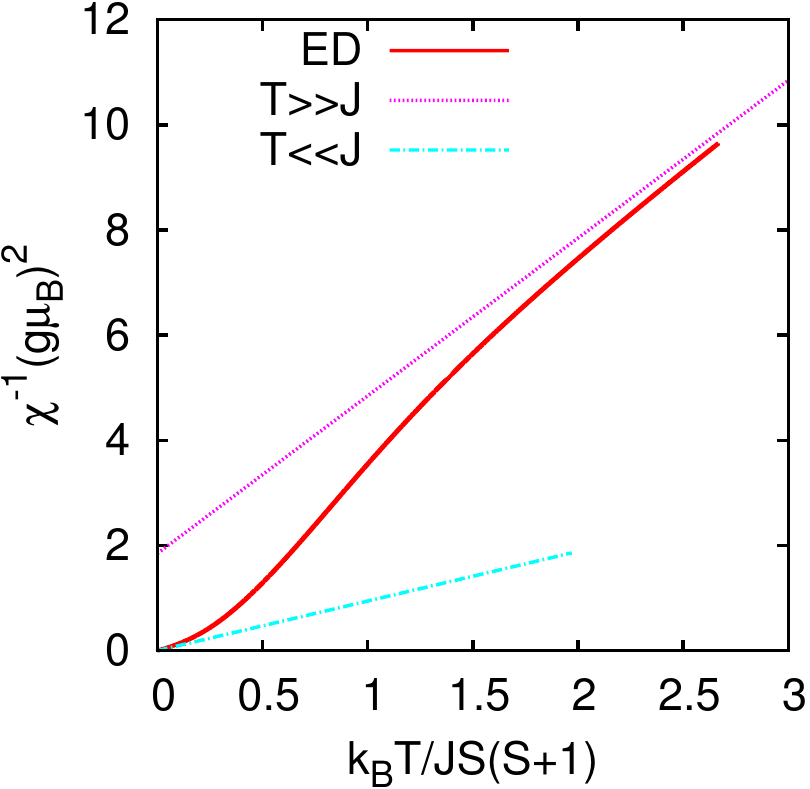}
\caption{Temperature dependence of inverse magnetic susceptibility per 
spin for the cluster shown in Fig.\ \ref{feoct}b, $S=0.5$. Dotted line shows 
the Curie-Weiss asymptotic for $T \gg J$, 
dot-dashed line shows superparamagnetic 
behavior with a super-spin $S_g=9S$ for $T \ll J$. }
\label{feoct2}
\end{figure}

For Fe-based double perovskites $T_{fe}$ is of the
order of the room temperature, as $J_1/k_B \sim 50$~K. From the graph 
shown in Fig.\ \ref{hte}a we see that in the range 
$ T_{fe}<T<T^{\ast}\approx 0.92J_1S(S+1)/k_B$, 
the magnetic susceptibility of the PFB2 phase exceeds the 
value for independent spins, $\chi (T) > \chi _{\mathrm{p}}(T)=S(S+1)/(3k_BT)$,
despite the antiferromagnetic character of the exchange interaction,
which suppress the magnetic response at high temperatures $T\gg J1$.
For comparison, the black thin solid line shows the susceptibility 
$\chi _{\mathrm{fcc}}(T)$
of 1:1 ordered PFB0 configuration, where Fe spins form a face centered
cubic lattice, and interact with $J_2=0.05J_1$. We see that 
$\chi _{\mathrm{fcc}}(T) < \chi _{\mathrm{p}}(T)$ at all temperatures
(see Appendix \ref{AppA}).

\subsection{Superparamagnetism}
A sample of a disordered double perovskite compound may contain some 
regions with PFB2 chemical order. In the ground state, such a region 
possesses the total
spin $S_g = (N_2-N_1)S$, where $N_1$, $N_2$ are the numbers of 
Fe1, and Fe2
sites in that region.\cite{Lieb62} 
In order to simulate the behavior of fragments of 
PFB2
configuration in a Fe-based double perovskite material, 
we show in 
Figs.~\ref{hte}b,~\ref{feoct} full exact-diagonalization data 
%for the $S=1/2$ case 
of thermodynamic
properties of clusters shown in Figs.~\ref{hte}b and \ref{feoct}(a-c).
Since we  have found that the dependence of the inverse susceptibility as a
function of normalized temperature $k_BT/J_1S(S+1)$ on the spin
value $S$ is weak (see Appendix \ref{AppA}),
the ED data for the simplest $S=1/2$ case can be
considered as represantative for higher values of $S$.  
The 7-site cluster shown in the  Fig.\ \ref{feoct}a contains one Fe1 
site
interacting with six Fe2 sites via $J_1$ exchange. 
This is a particular
case of the Heisenberg star model.\cite{Gaudin76,Richter94} 
For $T \gg
J_1S(S+1)/k_B$ the susceptibility per spin tends to 
the Curie-Weiss asymptotic 
$\chi _{\mathrm{CW}}=\chi _{\mathrm{p}}/[1+4S(S+1)J_1/7k_BT]$. 
In the opposite limit, the system
shows a \emph{super-paramagnetic} behavior,\cite{Bedanta09} 
i.e. it behaves as a single super-spin $S_g=5S$, and the 
susceptibility is 
$\chi_{\mathrm{SPM}}=S_g(S_g+1)\chi _{\mathrm{p}}/[(N_2+N_1)S(S+1)]$ 
(see Fig.~\ref{hte}b).
At temperatures $T\sim T_{fe}$ the
system exhibits a crossover between the two regimes. 
The susceptibility exceeds the 
independent-spin value for $T<T^{\ast}_1 \approx 0.74J_1S(S+1)/k_B$.
Similar results for a 13-site
cluster (Fig.~\ref{feoct}b) are shown in Fig.~\ref{feoct}d 
(see also Fig.~\ref{feoct2} in the Appendix).

In a real sample, an interaction between the regions of PFB2 
configurations
always exists. When the temperature becomes sufficiently low, the 
thermal and
interaction energies become comparable, and a collective state of 
super-spins
is formed. The behavior of two interacting PFB2 clusters 
(Fig.\ \ref{feoct}c) is shown in
Fig.\ \ref{feoct}d (lines $c_1$ and $c_2$).  
For temperatures $T \gg J_c$ the susceptibility behaves similar to
the non-interacting case. In particular, it exceeds the 
susceptibility of independent spins $\chi (T)> \chi _{\mathrm{p}}(T)$ at
 $T\lesssim T_{fe}$
and tends to the superparamagnetic behavior down to low temperature, 
where it exhibits a maximum (minimum at $\chi ^{-1}(T)$ curve). 
Below the maximum,
a singlet ground state of two interacting super-spins is formed. 
In reality, for large number of interacting clusters 
the disorder in the system  
favors a super-spin glass 
formation \cite{Bedanta09,Rotaru09,Kleemann10} at temperatures governed 
by the low energy scale $T<S(S+1)J_{2}/k_B$.

A characteristic feature of large spin formation in a system is a 
non-linearity of its magnetization curve $M(H)$, which results in the 
dependence of the susceptibility $\chi =M/H$ on the field value.
The Fig.\ \ref{hvsm} shows the $\chi ^{-1}(T)$ for the 
7-site cluster and a finite value of the magnetic field.
The susceptibility substantially deviates from the "theoretical" value
$\chi _{\mathrm{th}}= \partial M/\partial H (H=0)$ at low temperatures, 
and it does not diverge at $T \to 0$.
Note that we have considered here only isotropic 
Heisenberg interactions. The magnetic anisotropy, which is always 
present in real compounds \cite{Laguta13}
would transform the non-linear magnetization curves into 
narrow hysteresis loops.\cite{Bedanta09}

The  model of ferrimagnetism considered here can also be applied to 
PFN and PFT diluted by 
non-magnetic Ti and Zr ions.\cite{Sanchez11,Evans13,Sanchez13,Kumar09} 
As it was mentioned above, these systems 
show a sizable magnetic moment at room temperature in spite that the 
concentration of Fe ions was decreased up to 10\%. From a general point 
of view, the magnetic dilution will lead 
to the breaking of the infinite 
magnetic percolation clusters responsible for the long-range 
antiferromagnetic order 
as it was pointed out in Ref.~\onlinecite{Laguta13}. In a small magnetic cluster the 
probability of creation of a ferrimagnetic configuration of spins should 
be enhanced due to the limited number of interacting spins.  Moreover, the 
ferrimagnetic ordering can be realized on the edge 
of the (semi-)infinite antiferromagnetic
 cluster which size is of order of a few nanometers only because it is 
controlled by local fluctuations of the 1:1 composition between magnetic 
Fe and non-magnetic ions.\cite{Kleemann10}  Obviously, the "edge" 
effect becomes substantial with Ti and Zr doping.

A large room-temperature magnetic response is reported in many
wide-gap diluted magnetic semiconductors, such as GaN, ZnO, and
TiO$_2$ isovalently doped by transition 
metals,\cite{Dietl10,Janisch05,Ogale10}
"in which no ferromagnetism was expected at any temperature".\cite{Dietl10} 
We think that a solution of this puzzle may be a
formation of ferrimagnetic superstructure clusters similar to 
that we have considered here. Fig.\ 4 shows two examples of a planar
ferrimagnetic arrangement that may be formed by
magnetic impurities substituting for cations 
in zinc blend and wurtzit semiconductors. We see that these 
arrangements may form infinite two-dimensional sublattices, they also
may 
be transformed into tree-dimensional superstructures if they will be
connected by bridging spins having antiferromagnetic interactions 
with both planes,
the planes then will be ordered parallel.

\section{Conclusions.}
In summary, we 
have studied an example of a system of 
\emph{antiferromagnetically} interacting 
equal spins relevant for 
Fe-based double perovskite compounds, and having a \emph{ferrimagnetic} 
ground state.
We have  estimated  the transition temperature $T_{fe}$ for a 
macroscopic system,
and have argued that it can be close to  
room temperature. 
Such kind of ferrimagnetism
may be the origin of the room-temperature magnetism of
PFT/PZT and PFN/PZT systems.
For small clusters of the same structure we have shown that 
their magnetic 
susceptibility exceeds the susceptibility of independent spins at  
temperatures $T \lesssim T_{fe}$.
This gives a possible microscopic explanation for 
the still 
puzzling monotonous increase of the magnetic
susceptibility with decreasing 
temperature below N\'eel temperature, which is observed practically in all 
Fe-based double perovskites.
The ferrimagnetism of this kind 
may be responsible also for 
numerous observations of an unexpected large room-temperature magnetic 
response in 3$d$-metal based oxides.

\begin{acknowledgments} The authors thank M.\ D.\ Kuz'min and 
M.\ Mary\v{s}ko for very useful
discussions. The projects GACR 13-11473S, 
and NASc of Ukraine 07-02-14 
are acknowledged.
The exact diagonalization calculations were performed using 
J. Schulenburg's {\it spinpack}. 
\end{acknowledgments}

\appendix

\section{Details of numerical calculations}\label{AppA}

In the ideal 1:1 chemical order, Fe$^{3+}$ and the non-magnetic M$^{5+}$
ions alternate in the B position of the perovskite lattice ABO$_3$.
In this configuration (called PFB0 in Ref.~\onlinecite{Raevski12}), 
magnetic Fe$^{3+}$
ions form regular face centered cubic sublattice with
antiferromagnetic interaction $J_2$
between nearest spins in the sublattice.
The HTE results for the  PFB0 lattice are shown in 
the Fig.~\ref{pfb0}.
For such a lattice, a transition
into so called I-type antiferromagnetic order
(see insert of Fig.\ \ref{pfb0}) occurs at 
$T_I \approx  -\Theta _{CW,0}/5.76 \approx 0.69S(S+1)J_2/k_B$\cite{Pirnie66}, 
$\Theta _{CW,0}=  -4S(S+1)J_2/k_B$ being the paramagnetic Curie-Weiss 
temperature. Note that the whole curve $\chi^{-1}(T)$
lies {\em above} the Curie-Weiss  asymptotic (CW). This is a "normal" 
behavior when the antiferromagnetic interactions 
\emph{suppress} the magnetic response of a spin system.

For the PFB2 spin arrangement, the temperature dependence of the 
susceptibility for different spin values is shown in Fig.~\ref{pfb2}. 
For $S=1/2$-plot the tenth-order HTE\cite{Lohmann14} was used. 
A reasonable estimate of the temperature for the transition into
the ferrimagnetically ordered phase
$T_{fe}$ is given by that point where 
$\chi ^{-1}(T_{fe})=0$.
The values of $T_{fe}$ for different spins are given in 
the Table~\ref{tab:Tfe} of the main text. 

Fig.~\ref{feoctc} shows that the dependence of the  magnetic 
susceptibility on the spin value is very weak if we plot $\chi ^{-1}$ as the 
function of  $k_BT/JS(S+1)$. Thus, for the clusters shown in 
Figs.~\ref{feoct}b,c we may consider the simple $S=1/2$ case to be representative
for the other spin values, too.

The behavior of a larger cluster of the PFB2 configuration shown in 
Fig.~\ref{feoct}b is qualitatively similar to the previous cluster 
(see Fig.~\ref{feoct2}).
But now at low temperatures it behaves like a larger single 
spin $S_g=9S$.

 \bibliography{spmth} 

\begin{thebibliography}{53}
\bibitem [{rae()}]{raevtab}%
  \BibitemOpen
  \href@noop {} {}\bibinfo {note} {Note that, in this paper, the rows 2 and 4
  of the Table.II are interchanged compared with the Fig.3.}\BibitemShut
  {Stop}%
\end{thebibliography}
\end{document}